\documentclass[aps,twocolumn,showpacs,preprintnumbers,amsmath,amssymb,a4paper]{revtex4}

\usepackage{color}

\usepackage{colordvi}
\usepackage{amssymb}
\usepackage{amsmath}
\usepackage{epsf}
\usepackage{mathrsfs}
\usepackage{graphicx}
\usepackage{appendix}

\def\nn{\nonumber}
\def\be{\begin{equation}}
\def\ee{\end{equation}}
\def\bea{\begin{eqnarray}}
\def\eea{\end{eqnarray}}
\begin{document}

\title{Creating superfluid vortex rings in artificial magnetic fields }
\author{Rashi Sachdeva}
\email{rashi.sachdeva@oist.jp}
\affiliation{Quantum Systems Unit, Okinawa Institute of Science and Technology Graduate University, Okinawa 904-0495, Japan}

\author{Thomas Busch}
\affiliation{Quantum Systems Unit, Okinawa Institute of Science and Technology Graduate University, Okinawa 904-0495, Japan}

\date{\today}

\begin{abstract}

Artificial gauge fields are versatile tools that allow to influence the dynamics of ultracold atoms in Bose-Einstein condensates. Here we discuss a method of artificial gauge field generation stemming from the evanescent fields of the curved surface of an optical nanofibre. The exponential decay of the evanescent fields leads to  large gradients in the generalized Rabi frequency and therefore  to the presence of geometric vector and scalar potentials.  By solving the Gross-Pitaevskii equation in the presence of the artificial gauge fields originating from the fundamental HE$_{11}$ mode of the fibre,  we show that vortex rings can be created in a controlled manner. We also calculate the magnetic fields resulting from the higher order HE$_{21}$, TE$_{01}$, and TM$_{01}$ modes and compare them to the fundamental HE$_{11}$ mode. 
\end{abstract}


\maketitle
  
\section{Introduction}

Ultracold atomic condensates have emerged as suitable and flexible systems to study a variety of effects relating to condensed matter physics \cite{lewenstein_adv_phys, bloch_rmp}. These effects include many connected to the periodicity found in solid state systems by using gases trapped in optical lattices, but more recently also ones relating to the application of  gauge fields.

 Among them a particularly interesting one is the appearance of vortex structures in magnetic fields above a critical field strength, similar to the physics exhibited by type-II superconductor systems \cite{abrikosov}.  However, atomic Bose-Einstein condensates (BECs) are charge neutral, and hence real magnetic fields have no gauge-field like effects. Nevertheless, one can rely on methods that apply magnetic field to such neutral systems in an artificial manner. Many proposals to generate artificial gauge fields for BECs have been put forward in recent years, and several  have been successfully implemented \cite{lewenstein_adv_phys, gauge}. A conceptually convenient way is to rotate a BEC  \cite{coddington}, which mimics the Lorentz force experienced by a charged particle in a magnetic field, and shows the quantization of circulation in BECs by forming vortices and vortex lattices. Another way to generate artificial gauge fields is through Raman lasers \cite{spielman}, which allow to generate highly stable artificial magnetic fields of large amplitude. 

While vortex systems in BECs have been thoroughly studied, vortex rings, which are three dimensional structures with a closed loop core (i.e., a vortex line that loops back into itself) \cite{He_book}, have been harder to experimentally create and control. Experimental observation of vortex rings has been achieved dynamically, in superfluid helium \cite{He_vortex_rings} as well as for BEC systems through the decay of dark solitons in two-component BECs \cite{BEC_VR1}, direct density engineering \cite{BEC_VR2, BEC_VR3}, in the evolution of colliding symmetric defects \cite{BEC_VR4} and very recently in the time evolution of superfluid Fermi gases \cite{VR_zwierlein}. Theoretical proposals for the creation of vortex rings in a stationary state include interfering two component BECs \cite{BEC_VR5}, using a spatially dependent Feshbach resonance \cite{BEC_VR6}, and direct phase imprinting methods \cite{BEC_VR7}. For inhomogeneously trapped BECs, however, vortex ring structures are known to be unstable, which results in very short lifetimes and significant difficulties for experimental observation. The vortex rings either decay into elementary excitations by drifting towards the edge of the condensate \cite{BEC_VR5}  or annihilate within the condensate bulk. 

Motivated by the large interest in the study of vortex ring structures in BECs, we propose here a way to engineer artificial gauge fields for a BEC such that vortex rings form naturally. For this we study a system in which an atomic BEC is coupled to an optical nanofibre \cite{Vetsch} and show that the detailed control over the evanescent field outside the fibre allows to obtain control over the creation of stable vortex rings. In time-dependent gauge fields, this can also be used to study vortex ring dynamics in a controlled way.

Optical nanofibres have in recent years emerged as versatile tools for tailoring optical near field potentials that can interface with other quantum systems, as they offer controlled propagation of light inside and outside the fibre surface \cite{Yariv, Nieddu, Hakuta_qdots}. For this reason, experiments are currently carried out in many labs worldwide that explore the possibility of trapping and manipulating cold atomic gases using optical nanofibres \cite{Dowling, FLKien1, balykin2004, sile1, helical, tara, rauschenbeutel_2008njp, sile2, sile3}. Using two-color evanescent fields around nanofibres, an optical dipole trap for laser cooled atoms near to the fibre surface has already been realized \cite{Vetsch}, which proves that tapered optical fibres are excellent candidates for realizing versatile light matter interfaces. 

In this work, we consider the adiabatic motion of trapped Bose-Einstein condensed atoms around an optical nanofibre and show that the presence of the evanescent fields around the fibre realises interesting artificial gauge fields for the BEC system. Because of the large gradient in the generalized Rabi frequency, geometric vector and scalar potentials are created, which are related to Berry phases and have similar effects on the neutral atoms as the magnetic and electric fields have on charged systems \cite{Dalibard_RMP}. In particular we show that for the fundamental mode of the fibre the artificial magnetic field lines can go solely along the azimuthal direction and therefore allow for the creation of vortex rings. Similar studies to generate artificial gauge fields atop a flat surface, such as above a prism, have recently been presented as well \cite{Ev_AGF, Ev_AGF2}. 

The manuscript is organized as follows. In Section \ref{Sec:Background}, we briefly discuss the background for our work by first reviewing the general model for the adiabatic motion of atoms in an external electromagnetic field and describing how an artificial vector potential is generated outside of a dielectric surface.
We also review the explicit forms of the evanescent fields. In Section \ref{Sec:AGF} we calculate the expressions for the effective magnetic field profiles resulting from the combination of different polarisations of fundamental HE$_{11}$ modes of the fibre and describe their effect on Bose-condensed atoms trapped around the fibre in Section \ref{Sec:GPE_results}. In Section \ref{Sec:HOM} we discuss different  magnetic field profiles that can be obtained from higher order modes, and in Section \ref{Sec:Summary} we conclude. 

\section{Background} \label{Sec:Background}
\subsection{Adiabatic motion of atoms in evanescent fields} 
\label{Sec:AdiabaticMotion}

Let us start by considering a two level atom at position $\mathbf{r}$, which interacts with an external laser field \cite{CCT:2011}.  Within the rotating wave approximation its eigenstates are called dressed states and are given by  
\begin{align}
 	|\Psi_1(\mathbf{r})\rangle &=\left(\begin{array}{l}
																\cos(\Phi(\mathbf{r})/2) \\
																\sin(\Phi(\mathbf{r})/2)~e^{i\phi(z)} 
																\end{array} \right), \\
 	|\Psi_2(\mathbf{r})\rangle &=\left( \begin{array}{l}
																-\sin(\Phi(\mathbf{r})/2)~e^{-i\phi(z)}\\
																\phantom{-}\cos(\Phi(\mathbf{r})/2) 
 																\end{array} \right) .
\end{align}
Here $\phi(z)$ is  the running phase of the optical field and $\Phi(\mathbf{r})=\text{arctan}(|\kappa(\mathbf{r})|/\Delta)$, where 
\begin{equation}
	\kappa(\mathbf{r})=\mathbf{d}.\mathbf{E(r)}/\hbar
\end{equation} 
is the system's Rabi frequency with $\mathbf{d}$ and $\mathbf{E(r)}$ being the atomic dipole moment and the electric field vector, respectively. The detuning of the light field from the resonance frequency $\omega_0$  is given by $\Delta=\omega_0-\omega$. The states are split in energy by $\epsilon_1(\mathbf{r})-\epsilon_2(\mathbf{r})=\hbar\Omega(\mathbf{r})$, where $\Omega(\mathbf{r})=\sqrt{\Delta^2+|\kappa(\mathbf{r})|^2}$ is the generalised Rabi frequency.   

Assuming that the atom is initially prepared in state $|\Psi_1(\mathbf{r})\rangle $ and moves adiabatically in the external light field, its internal state will also adiabatically follow the dressed state. This leads to the appearance of a geometrical Berry phase, and hence a vector potential of the form
\begin{align}
		\mathbf{A}&=i\hbar\langle \Psi_1|\nabla\Psi_1\rangle,\\
		 &= \frac{\hbar}{2}~[\cos(\Phi(\mathbf{r}))-1]~\nabla\phi(\mathbf{r}). 
	\label{vecpot} 
\end{align}

This represents an artificial gauge potential which is geometric in nature, since it arises from the spatial variation of the dressed state. The system can therefore mimic the dynamics of a charged particle in the presence of magnetic field, 
given by $\mathbf{B}=\nabla\times \mathbf{A}$. 

 In the following, we will use the properties of evanescent fields outside of optical nanofibres to generate artificial magnetic fields for adiabatically moving ultracold atoms. We will show that these fields can have different profiles, depending on the mode characteristic of the light travelling through the fibre. 

If we assume that the field travels freely along the fibre, we can choose $\phi(z)=k_0 nz$, with $k_0$ as the wave number and $n$ as the refractive index, and straightforwardly calculate the vector potential as
\begin{equation} 
\mathbf{A}(\mathbf{r})=-\hat{z} \frac{\hbar k_0 n }{2}\left[1-\frac{1}{\sqrt{1+\big(\frac{|\mathbf{d}\cdot\mathbf{E}|}{\hbar\Delta}\big)^2}}\right],
\label{vecpot2}
\end{equation}
from which the artificial magnetic field follows as

\bea \mathbf{B}(\mathbf{r})&=&\frac{\hbar k_0 n}{4} \frac{(d/\hbar\Delta)^2}{\left[1+\left\{\frac{|\mathbf{d}\cdot\mathbf{E}|}{\hbar\Delta}\right\}^2\right]^\frac{3}{2}}\left[\hat{\varphi}~\frac{\partial}{\partial r}|\mathbf{E}|^{2} -\hat{r}~\frac{1}{r}\frac{\partial}{\partial \varphi}|\mathbf{E}|^{2}\right].\nn\\
\label{magfield}\eea
Since we have evaluated this expression in cylindrical polar coordinates, one can immediately see that  the resulting B-field has components pointing along the $\hat{\varphi}$ and the $\hat{r}$ direction. 
While evanescent field modes have inevitably an $r$ dependence, for modes of the nanofibre which have no azimuthal dependence (for example the ones with circular polarisation), only the magnetic $\hat{\varphi}$ component exists. This is the basis for the ability to generate vortex rings around the fibre.

As can be seen from Eq.~\eqref{vecpot}, atoms interacting with fields that have large gradients are subject to stronger artificial gauge fields. Evanescent fields outside of optical nanofibres are known to have very large field gradients, and hence these systems are of experimental interest. Since the gradients also depend on the refractive index and the diameter of the fibre, as well as the parameters of the input light field, a large number of valuable control parameters exist with which the strength and spatial structure of the artificial magnetic fields can be changed. 

\subsection{ Form of the evanescent fields}\label{Sec:Atomtrap}

Optical nanofibres can be thought of as consisting of an extremely thin cylindrical silica core and an infinite vacuum clad. They can be created by heating and pulling a standard commercial grade optical fibre so that its waist diameter reduces from a few hundred micrometers to a few hundred nanometers \cite{Yariv, sile1,mazur_2003}. Since the fibre diameter is smaller than the wavelength of the input light, a major fraction of power propagates outside the surface in the form of an evanescent field. 

Trapping of atoms around the fibre can be achieved using a setup that relies on two evanescent fields \cite{balykin2004, Dowling, FLKien1,helical,tara,Vetsch}. The first field is red-detuned with respect to atomic transition frequency and provides a potential that attracts atoms towards the fibre. The second field is blue-detuned with respect to atomic transition frequency, leading to a potential that repulses the atoms from the surface. Since both fields have different evanescent decay lengths, it is possible to create a potential minimum in the radial direction at a finite distance ($\sim 200$ nm) away from the fibre surface. 

In this work we will explicitly consider the effects of light propagating in the fundamental HE$_{11}$ mode of the nanofibre, where the frequency, the free space wave number and the wavelength are denoted by $\omega$, $k_0=\omega/c$ and $\lambda=2\pi/k_0$, respectively. To ensure that only this fundamental mode propagates in the fibre, the single mode condition $V=k_0a\sqrt{n_1^2-n_2^2}<V_c\approx 2.405$ needs to be fulfilled, where $a$ is the radius of the fibre and  $n_1$ and $n_2$ are the refractive indices inside and outside of the fibre. This can be easily achieved for typical nanofibre diameters.

For a circularly polarised light field, the components of the electric field vector for the fundamental HE$_{11}$ mode outside the fibre are given by
\begin{align}
		E_r&=iA[(1-s)K_0(qr)+(1+s)K_2(qr)]~e^{i(\omega t-\beta z)},\nn\\
        E_\varphi &=-A[(1-s)K_0(qr)-(1+s)K_2(qr)]~e^{i(\omega t-\beta z)},\nn\\
        E_z&=2A(q/\beta)K_1(qr)~e^{i(\omega t-\beta z)},
        \label{ecircular}
\end{align}
where $s$ is a dimensionless parameter given by
\begin{equation}
	s=\frac{1/h^2a^2+1/q^2a^2}{J_{1}'(ha)/ha ~J_{1}(ha)+K_{1}'(qa)/qa ~K_{1}(qa)}. 
	\label{sparameter}
\end{equation} 
The normalisation constant $A$ is defined as
\begin{equation} 
		A=\frac{\beta}{2q}\frac{J_1(ha)/K_1(qa)}{\sqrt{2\pi a^2(n_1^2 N_1+n_2^2 N_2)}}, 
		\label{Anorm}
\end{equation}
where
\begin{align} 
		N_1=& \frac{\beta^2}{4h^2}\bigg[ (1-s)^2[J_0^2(ha)+J_1^2(ha)]\nn\\
				& +(1+s)^2[J_2^2(ha)-J_1(ha)J_3(ha)]\bigg]\nn\\
				& +\frac{1}{2}[J_1^2(ha)-J_0(ha)J_2(ha)], \nn\\
		N_2=&\frac{J_1^2(ha)}{2K_1^2(qa)}\bigg\{ \frac{\beta^2}{4q^2}\bigg[ (1-s)^2[K_1^2(qa)-K_0^2(qa)]\nn\\
				& -(1+s)^2[K_2^2(qa)-K_1(qa)K_3(qa)]\bigg]\nn\\
				& -K_1^2(qa)+K_0(qa)K_2(qa) \bigg\}.
\end{align}
In above expressions, $J_m(x)$ and $K_m(x)$ are Bessel functions of the first kind, and modified Bessel functions of the second kind, respectively, and $\beta$ is the longitudinal propagation constant for the fibre's fundamental mode. The parameter  $q=\sqrt{\beta^2-n_2^2k_{0}^{2}}$  characterizes the decay of the field outside the nanofibre and $h=\sqrt{n_1^2k_{0}^{2}-\beta^2}$.   

When the input light field is linearly polarised, the components of the electric field vector of the evanescent field are given by
\begin{align} 
	 E_x=&\sqrt{2}A[(1-s)K_0(qr) ~\text{cos}~\varphi_0\nn\\
			& +(1+s)K_2(qr)~\text{cos}(2\varphi-\varphi_0)]~e^{i(\omega t-\beta z)},\nn\\
	E_y=&\sqrt{2}A[(1-s)K_0(qr)~\text{sin}~\varphi_0\nn\\
			& +(1+s)K_2(qr)~\text{sin}(2\varphi-\varphi_0)]~e^{i(\omega t-\beta z)},\nn\\
	E_z=&2\sqrt{2}iA(q/\beta)K_1(qr)~\text{cos}~(\varphi-\varphi_0)~e^{i(\omega t-\beta z)}.
	\label{elinear}
\end{align}
Here the angle $\varphi_0$ determines the orientation of the polarization, with $\varphi_0=0$ and $\pi/2$ being aligned along the $x$ and $y$ axes, respectively.
From these expressions one can see that for circularly polarised light fields, the atoms can be trapped in a cylindrical shell which surrounds the nanofibre. However, when either one or both of the input light fields (red and blue detuned) are linearly polarised, the trapping potential possesses minima at specific spatial points in the transverse plane of the optical fibre. 

\section{Artificial gauge fields for cold atoms trapped outside the nanofibre } \label{Sec:AGF}

To calculate the gauge fields stemming from the evanescent fields, one can see from Eq.~\eqref{magfield} that smaller detunings lead to larger fields. However, smaller detunings also lead to higher scattering rates and therefore higher losses \cite{FLKien1}. To avoid the latter, detunings used for atom traps are usually chosen to be quite large (order of THz), to ensure low scattering rates, giving coherence times of $\sim 50$ ms and trap lifetimes of up to $\sim 100$ s \cite{Vetsch}. This, however, leads to unobservably small gauge fields.

To overcome this limitation, an alternative arrangement  was recently suggested by M.~Mochol  and K.~Sacha \cite{Ev_AGF},  which does not depend explicitly on the detuning of the input light field. This scheme can be implemented for multilevel alkali-metal atoms such as $^{87}$Rb which are a commonly used species in cold atom experiments, and takes advantage of the (quasi) degeneracy of the electronic ground state level.  In the dressed state picture dark and bright states exist, which are linear combinations of the degenerate ground states, and which have negligible contributions from the excited state. The required coupling 
uses two light fields, the first propagating inside the fibre, and the second  propagating outside and parallel to the fibre surface. Both beams are assumed to have the same wavevector, $k_1\approx k_2=k_0$, and their respective Rabi frequencies are given by $\kappa_{1}(r,\varphi,z)$ and $\kappa_{2}(z)$. The values are chosen such that they can induce Raman transitions between two degenerate internal states of the atoms
and we assume that the atoms follow adiabatically the coupled dressed state given by
\begin{equation} 
	|D_1\rangle = \frac{|1\rangle+\xi |2\rangle}{\sqrt{1+|\xi|^2}}.
\end{equation}
Here 
\begin{align}
	\xi=&-\frac{\kappa_1^\ast}{\kappa_2^\ast}=-\frac{|\mathbf{d_1}\cdot\mathbf{E_1}|}{|\mathbf{d_2}\cdot\mathbf{E_2}|}\nn\\
	     =&-\tilde{s}(d_rE_r+d_{\varphi}E_{\varphi}+d_zE_z)e^{-ik_0(n_1+1)z}
\end{align}
with $\mathbf{d_1}=d_{01}(d_r\hat{r}+d_{\varphi}\hat{\varphi}+d_z\hat{z})$, and $\mathbf{E_1}=E_{01}(E_r\hat{r}+E_{\varphi}\hat{\varphi}+E_z\hat{z})$, hence $\tilde{s}=d_{01}E_{01}/|\mathbf{d_2}\cdot\mathbf{E_2}|$.  Using Eq.~\eqref{vecpot}, one can determine the effective vector potential and hence the magnetic field as
\begin{align} 
	\mathbf{A}(\mathbf{r})=&-\hat{z}\hbar k_0(n_1+1)\tilde{s}^2 \frac{|d_rE_r+d_{\varphi}E_{\varphi}+d_zE_z|^2}{1+\tilde{s}^2|d_rE_r+d_{\varphi}E_{\varphi}+d_zE_z|^2},\nn\\ \label{vecpotalt}\\
 	\mathbf{B}(\mathbf{r})=&\frac{\hbar k_0 \tilde{s}^2(n_1+1)}{(1+\tilde{s}^2|d_rE_r+d_{\varphi}E_{\varphi}+d_zE_z|^2)^2}\nn\\
 						& \times\bigg[\hat{\varphi}~\frac{\partial}{\partial r}|d_rE_r+d_{\varphi}E_{\varphi}+d_zE_z|^2\nn\\
						& -\hat{r}~\frac{1}{r}\frac{\partial}{\partial \varphi}|d_rE_r+d_{\varphi}E_{\varphi}+d_zE_z|^2\bigg]\label{magfieldalt}.	
\end{align}
One can see that this expression is independent of the detuning  and a significant magnetic field can be achieved by making the parameter $\tilde{s}$  at least of order unity or greater, which is experimentally achievable.

\begin{figure}[tb] 
	\includegraphics[width=1.0\linewidth]{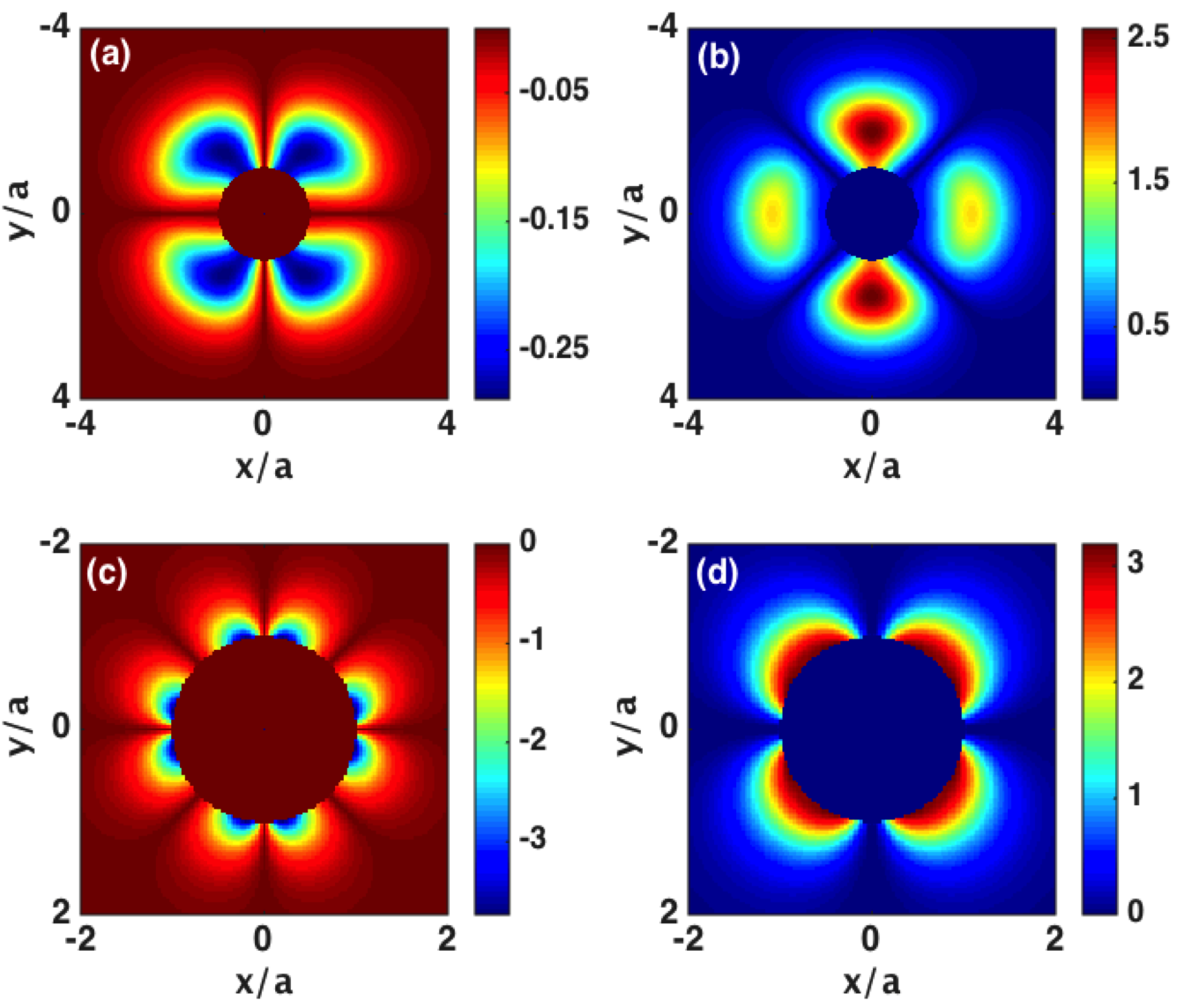} 
\caption{Magnetic fields strength $B(x,y)$ stemming from a linearly polarised HE$_{11}$ mode in units of $B_0=\hbar k_0^{2}/2$. (a)~$\hat{r}$ and (b)~$\hat{\varphi}$ components for polarisation in the $x$ direction, (c)~$\hat{r}$ and (d)~$\hat{\varphi}$ components for polarisation along $y$ . The wavelength and power of the blue detuned light field are chosen as $\lambda_B=700$~nm and $P_B= 30$~mW, and the components of the dipole moment are $d_r=1, d_{\varphi}=0$ and $d_z=0$. Note that the axes in (c) and (d) are adjusted.
 \label{LP_Br_Bphi} } 
\end{figure}

Again we note that the expression for the magnetic field has components along the $\hat{r}$ and the $\hat{\varphi}$ directions and 
in Fig.~\ref{LP_Br_Bphi} we show the spatial distribution of the different components for a blue detuned light field that is linearly polarised along the $x$ ($\varphi_0=0$) and the $y$ ($\varphi_0=\pi/2$) direction (cf.~Eq.~~\eqref{ecircular}). As before, for circularly polarised fields no radial component exists.

\begin{figure}[tb] 
	\includegraphics[width=1.0\linewidth]{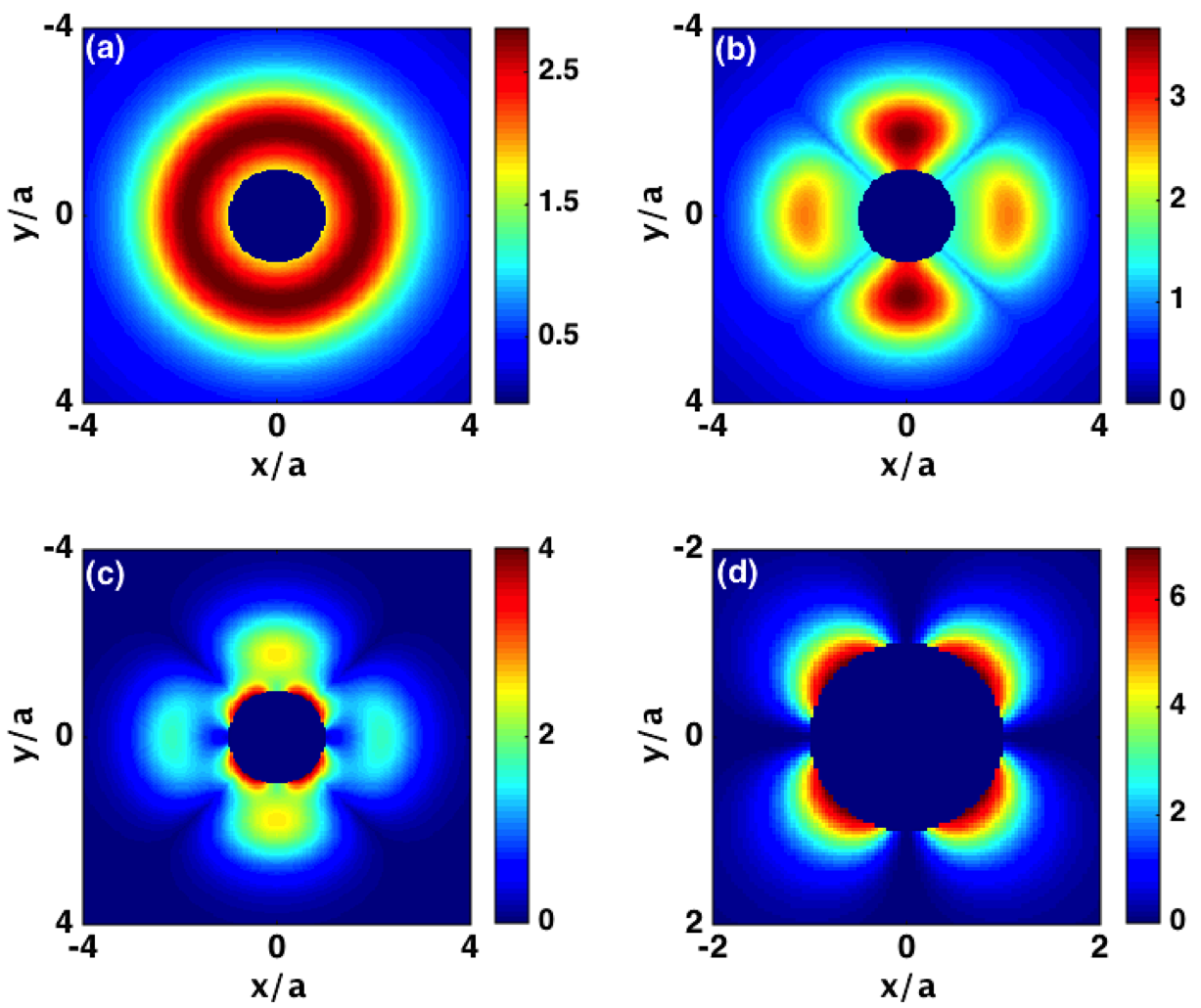} 
\caption{Magnetic field strength $B(x,y)$, in units of $B_0=\hbar k_0^{2}/2$, for atoms trapped outside a fibre of radius $a=200$~nm. The wavelength and power of blue and red detuned light fields are $\lambda_B=700$~nm, $P_B= 30$~mW, and $\lambda_R=1060$~nm, $P_R= 20$~mW. The components of the dipole moment are chosen to be $d_r=1, d_{\varphi}=0$ and $d_z=0$. The polarisations of the input light fields are (a) both circular, (b) red circular and blue linear along $x$, (c) red linear along $y$ and blue linear along $x$, (d) red linear along $y$ and blue linear along $y$. Note that the axes in (d) are adjusted. 
 \label{Bfield_pol} } 
\end{figure}

For traps relying on two color light fields, the resulting magnetic field strength profiles in the transverse $xy$- plane  for different combinations of the polarisations states of the input fields are shown in Fig.~\ref{Bfield_pol}. They are based on the assumption that the two beams required for the trapping are propagating through the fibre in the fundamental HE$_{11}$ mode and for each a second beam to overcome the dependence on the detuning is  added. The trapping wavelengths and powers chosen are compatible with trapping Cs atoms in a deep optical potential outside the fibre, but this method for artificial magnetic fields can be used for many atomic species by appropriately adapting  the trapping wavelengths and the nanofibre diameter. 

One can see that when both of the input light fields are circularly polarised (Fig.~\ref{Bfield_pol}(a)), the resulting magnetic field is uniformly distributed around the fibre. However, when either one or both of the input light fields are linearly polarised, the azimuthal symmetry is broken and the magnetic field profiles become non-uniform around the fibre, as shown in Figs.~\ref{Bfield_pol}(b)-(d). This simple example already demonstrates that the magnetic field profiles outside the fibre can in principle be continuously and time-dependently tuned by controlling the polarisation state of the two input light fields. From now onwards, we will focus on the azimuthally symmetric situation and therefore consider both light fields to be circularly polarised modes as given by Eq.~\eqref{ecircular}.

\begin{figure}[tb] 
\includegraphics[width=1.0\linewidth]{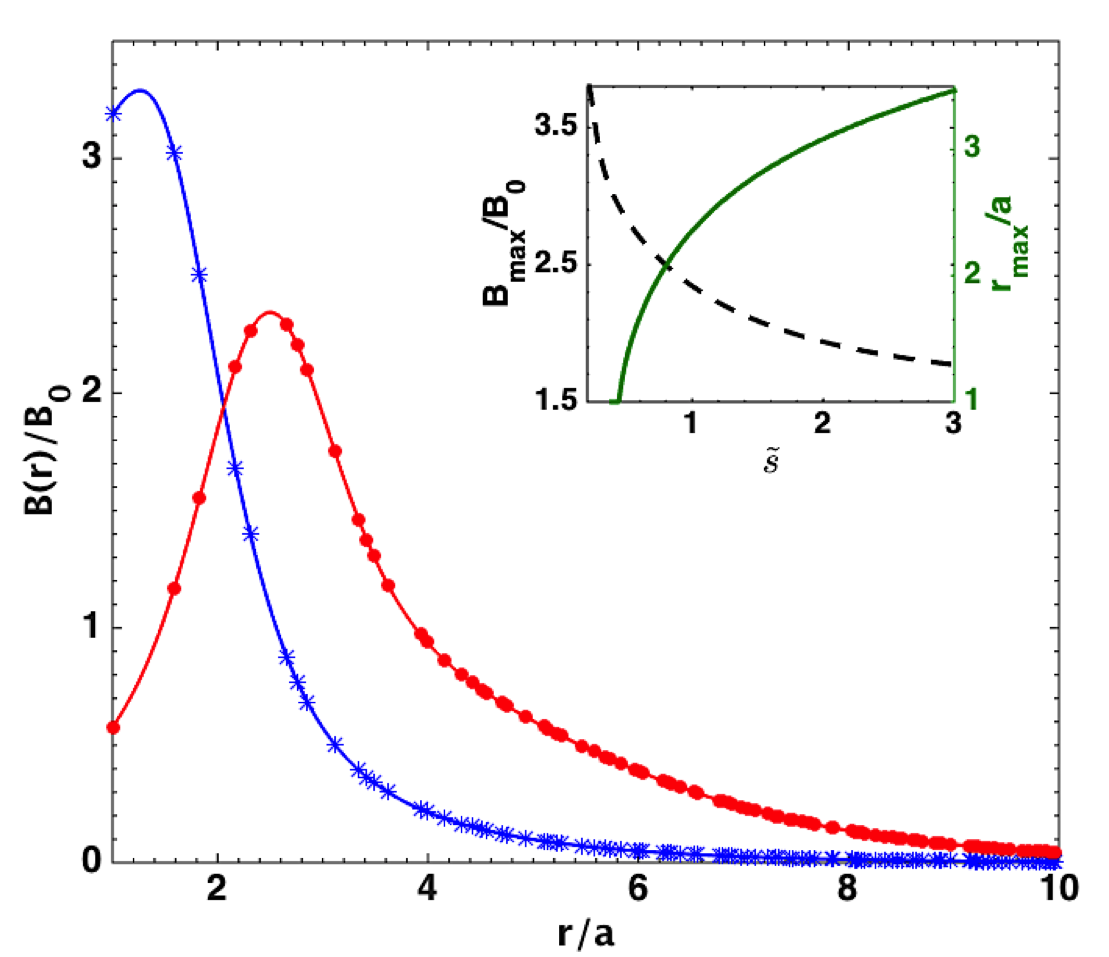} 
\caption{Magnetic field $B(r)$ scaled in units of $B_0=\hbar k_0^{2}/2$ outside a fibre with radius $a=200$~nm for different values of the parameter $\tilde{s}$. Blue (with asterisk) curve: $\tilde{s}=0.3$ ; Red (with dots) curve: $\tilde{s}=1$. For both values the components of the dipole moment are chosen to be $d_r=1, d_{\varphi}=0$ and $d_z=0$.  The inset (left axis) shows the change of the magnitude of the field maximum ($B_\text{max}/B_{0}$) as a function of $\tilde{s}$  (black dashed line). The right axis of the inset shows the location $r_\text{max}$ of maximum of the magnetic field (green solid line).  The wavelengths and powers of the blue and red detuned light fields are the same as in Fig.~\ref{Bfield_pol}. 
\label{Baniso} } 
\end{figure}

From Fig.~\ref{Bfield_pol}(a) it can also be seen that the magnetic field possess a maximum at a finite distance away from the fibre surface and decreases rapidly beyond that. The exact position and value of this maximum is a function of the parameter $\tilde{s}$ of the two input light fields and of the dipole moment components $d_r, d_{\varphi},d_z$.   Shifting the maximum of the magnetic field away from the fibre to achieve a better overlap with an atomic cloud, however, requires a compromise with the maximum value of the magnetic field, which reduces with increasing distance from the fibre, see Fig.~\ref{Baniso}. 
The inset of this figure shows the magnitude, $B_\text{max}$, and the position, $r_\text{max}$, of the maximum of the magnetic field as a function of parameter $\tilde s$. One can see that with increasing values of $\tilde s$, the maximum moves further away from the fibre surface, but decreases in magnitude. In the next section we show what the effect of these artificial magnetic fields is on a typical BEC trapped around the nanofibre.

\section{Bose Einstein condensates in artificial magnetic fields around the nanofibre}\label{Sec:GPE_results}

The fact that the trapping potentials and the artificial magnetic fields arising from the circularly polarised HE$_{11}$ modes are azimuthally symmetric allows us to restrict our calculations for the effects of the gauge fields on the BEC to the $xz$ plane, which significantly reduces the required numerical resources. Furthermore, to constrain the extent of the condensate in the $z$-direction, and since the exact shape of the trapping potential does not play a role in describing the physics, we will approximate it by a harmonic shape, choosing the harmonic potential in the radial direction to have a minimum at the same location as that of the two color trapping potential. 

\begin{figure}[t] 
\begin{center} 
\includegraphics[width=1.0\linewidth]{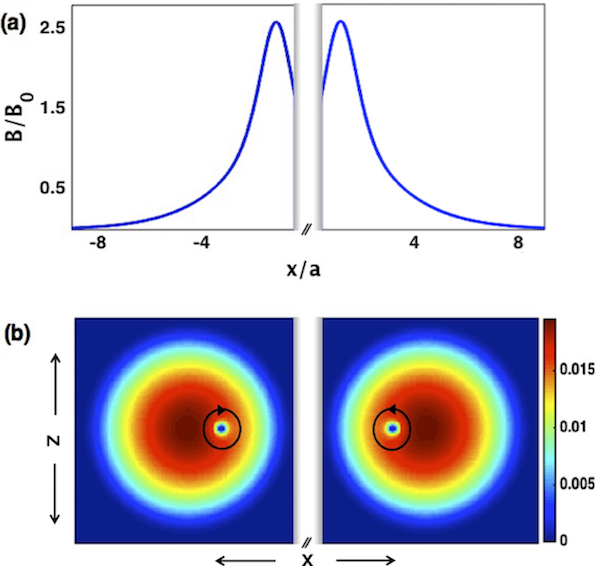} 
\end{center} 
\caption{(a) Magnetic field strength as a function of distance from the nanofibre surface. (b) Density profile for a BEC trapped in the harmonic potential on left and right side outside an optical nanofibre. The vortices visible are the result of the presence of the artificial magnetic field created by the evanescent field outside the fibre. The artificial magnetic field used in the calculations correspond to the laser parameters used in Fig.~\ref{Bfield_pol} (a) with $\tilde{s}=0.7$ and $d_r=1, d_{\phi}=0, d_z=0$.}
 \label{vortices1} 
\end{figure}

\begin{figure}[t] 
\begin{center} 
\includegraphics[width=01.0\linewidth]{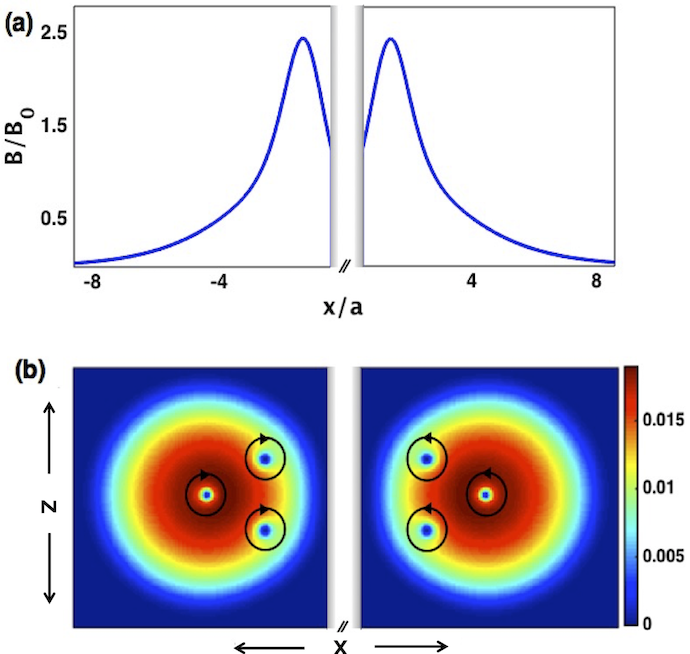} 
\end{center} 
\caption{Same as Fig.~\ref{vortices1}, but for $\tilde{s}=0.85$.
One can see that the broader magnetic field distribution leads to the appearance of multiple vortices. 
}  
\label{vortices2} 
\end{figure}

 We use the total artificial magnetic field created by the red and blue detuned input light fields and, assuming that the condensate dynamics can be described within the mean field approximation,
 solve the time-independent Gross Pitaevskii equation given by 
\begin{equation}  
	-\frac{1}{2}\left(\nabla+i\mathbf{A}\right)^{2}~\psi +V_{\text{trap}}~\psi+g|\psi|^2~\psi=\mu\psi,
	\label{gpe}
\end{equation}
where $V_{\text{trap}}$ is the harmonic trapping potential. The interaction between the atoms is characterised by $g$ and $\mu$ is the chemical potential of the system. The ground state of this equation can be easily found using a fft/split-operator method in imaginary time and results for two different strengths of the artificial magnetic field are shown in  Figs.~\ref{vortices1} and \ref{vortices2}. 

The magnetic field strength in the radial direction away from the fibre for $\tilde s=0.7$ is shown in the upper part of Fig.~\ref{vortices1} and the lower part  shows the density profile of the condensate on the left and right hand side of the fibre. The surface of the fibre is indicated in the middle of both plots. One can see that a single vortex appears on each side of the fibre, close to the position where the gauge field is maximal.

Calculating the circulation of the two vortices shows that they have equal and opposite values,
which is due to the fact that the magnetic field lines circulate around the fibre in the azimuthal direction. They therefore act perpendicularly to the $xz$- plane on both sides, but in opposite directions, which results in a vortex on the left hand side of the fibre and an anti-vortex on the right hand side. Restoring the azimuthal symmetry, the two vortices become slices through a vortex ring that is created around the fibre. The specific geometry of a condensate trapped around a nanofibre therefore allows to create vortex rings in a deterministic manner. 

Increasing the value of the parameter $\tilde s$ leads to a decrease of the magnitude of the artificial gauge field, but also to an increase in the width of the magnetic field profile. This increases the overlap with the trapped condensate and one can see in Fig.~\ref{vortices2} that this results in the generation of multiple vortex rings around the fibre. It is worth noting that these solutions are only stable in the presence of the gauge field, i.e.~when light is propagating through the fibre. However, the field inside the fibre can be changed time-dependently 
and typical timescales required for changing the detuning, the power or the polarization of the input light fields are of the order of milli- to micro-seconds, which is a lot shorter than the typical life times of atomic BECs.
It is worth noting that a particular interesting situation is the one where the fields are switched off after the vortex rings have formed. The topologically stable rings can then evolve freely inside the condensate, which would allow to study the dynamics of vortex ring interactions starting from a well-defined initial state. To account for all possible effects, which include oscillations along the vortex ring as well as reconnections, such work needs to be carried out in fully three-dimensional simulations \cite{James:17}.

\section{Artificial magnetic field due to evanescent fields from higher order modes} \label{Sec:HOM}

\begin{figure}[t] 
\begin{center} 
\includegraphics[width=1\linewidth]{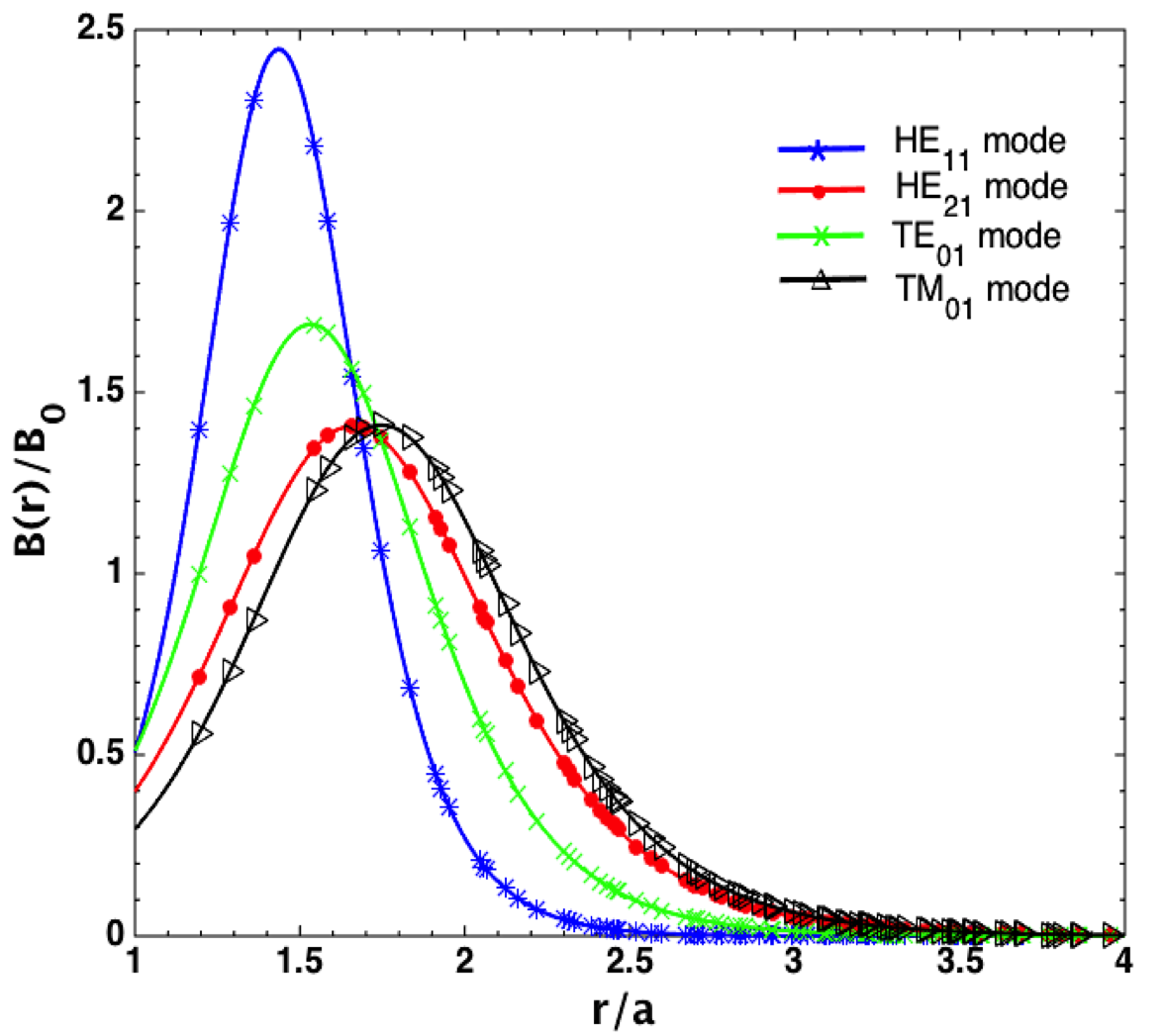} 
\end{center} 
\caption{Magnetic fields $B(r)$, scaled in units of $B_0=\hbar k_0^{2}/2$, for the higher order modes HE$_{21}$ (red line with dots), TE$_{01}$ (green line with cross), TM$_{01}$ (black line with triangles), and compared to the fundamental HE$_{11}$ mode (blue line with asterisk).  The wavelength of blue detuned light field is $\lambda_B=780$~nm and dipole moment components are $d_r=1, d_{\phi}=1, d_z=1$ respectively, with parameter $\tilde s=5$. The fibre radius is chosen to $a=400$~nm. \label{HOM}} 
\end{figure}

So far we have focussed on the effects of the fundamental HE$_{11}$ mode, which can be separated from the remaining optical modes by a proper cut-off condition. However, higher-order mode transmission in nanofibres has recently been achieved \cite{sile2}, which has numerous applications, for example in engineering new trapping geometries for atoms based on the different evanescent field shapes \cite{helical, tara, rauschenbeutel_2008njp}. These higher order modes bring with them additional degrees of freedom, which allow for more flexible artificial magnetic field profiles and we show in Fig.~\ref{HOM} the respective profiles resulting from the three higher-order modes TE$_{01}$, TM$_{01}$ and HE$_{21}$, which are the ones closest to the fundamental mode HE$_{11}$. The explicit expressions for their evanescent fields 
are given in the Appendix.
In order to allow the higher order modes to travel through the nanofibre, a larger fibre radius is required and we focus on a single, blue detuned input light field of wavelength $ \lambda_B=780$~nm and power $P_B=30$~mW for a fibre of radius $a=400$~nm. From Fig.~\ref{HOM} one can see that the magnitude is highest for the HE$_{11}$ mode and is decreased for the TE$_{01}$,  TM$_{01}$ and HE$_{21}$ modes. At the same time, the width of these higher order modes increases, which is consistent with the fact that their evanescent fields have larger decay lengths. As before, these magnetic field profiles can also be tuned by changing the parameter $\tilde{s}$ and the dipole moment components $d_{r}$, $d_{\varphi}$ and $d_{z}$. 

Furthermore, it is in principle possible to interfere different order modes and thereby engineer non-trivial evanescent field profiles  \cite{helical,tara}, which in turn will lead to complex magnetic field geometries and potentially new structures inside the BEC. 
%This is due to the different evanescent decay lengths, as well as shape of the electric field components of the different modes propagating in the fibre. Higher order modes thus provides a rich platform to design magnetic field profiles and we explore it in detail in one of our future works 
\cite{James:17}.

\section{Summary and outlook}\label{Sec:Summary}

In this work we have described the artificial magnetic fields stemming from the evanescent fields of an optical nanofibre and their effects on cold atoms trapped around such fibres. The strong gradient of the fields combined with the adiabatic motion of the atoms leads to a  geometrical Berry phases that can be represented by vector and scalar potentials experienced by the atoms. We have shown that the vector potential can lead to a magnetic field that has components in the radial and the azimuthal direction, and that the component in the radial direction can be removed.
If a Bose Einstein condensate is placed in such an evanescent field, the synthetic magnetic field can induce vorticity in the condensate, and due to the geometry of the setup, this can lead to the controlled formation of vortex rings. 

While in this work we have only examined the stationary states inside the artificial magnetic field, it is of larger interest to also consider possible dynamical scenarios. As the magnetic field is purely based on optical fields, and since these can be changed in a time-dependent manner, the system presented above suggests itself for dynamical studies as well.  Fast changes of the detuning, the power or the polarization of the input light fields would allow to study quenched systems, whereas a controlled reduction of the field strength would allow to 'unpin' the vortex rings and study free vortex ring dynamics starting from a well defined initial state. In addition to this a large number of structures resulting from modes that do not have azimuthal symmetry exist and can be characterised. However, all of these studies will require a fully three-dimensional treatment and we are currently preparing for this.

\section*{Acknowledgements}  The authors are grateful to Y.~Zhang, T.~Hennessey, C.~Madiah, F.~Le Kien and T.~Ray for fruitful discussions and suggestions. This work was supported by the Okinawa Institute of Science and Technology Graduate University.

\appendix*
\section{Higher order modes}
\label{Appendix}

In this Appendix we give the explicit expressions for the evanescent fields of the higher order modes used in Section \ref{Sec:HOM}.

\subsection{HE$_{21}$ mode}

The electric field components outside the fibre core for the HE$_{21}$ mode  are given by
\begin{align}
E_{r}&=iA_{21}[(1-u)K_1(qr)+(1+u)K_3(qr)]~e^{i(\omega t-\beta z)},\nn\\
E_{\varphi}&=-A_{21}[(1-u)K_1(qr)-(1+u)K_3(qr)]~e^{i(\omega t-\beta z)},\nn\\
E_{z}&=2A_{21}(q/\beta)K_2(qr)~e^{i(\omega t-\beta z)},\label{ecircular_HE21}
\end{align}
where $u$ is the dimensionless parameter
\begin{equation} 
    u=\frac{2(1/h^2a^2+1/q^2a^2)}{J_{2}'(ha)/ha ~J_{2}(ha)+K_{2}'(qa)/qa ~K_{2}(ha)}. \label{u_HE21}
\end{equation}
The normalisation constant $A_{21}$ is defined as
\be A=\frac{\beta}{2q}\frac{J_2(ha)/K_2(ha)}{\sqrt{\pi} a\sqrt{(n_1^2 R_1+n_2^2 R_2)}}, \label{A_21}\ee
where
\begin{align} 
R_1 &= \frac{\beta^2}{2h^2}\bigg[ (1-u)^2[J_1^2(ha)-J_0(ha)J_2(ha)]\nn\\
&  +(1+u)^2[J_3^2(ha)-J_2(ha)J_4(ha)]\bigg]\nn\\
&  +[J_2^2(ha)-J_1(ha)J_3(ha)], \nn\\
R_2 &=\frac{J_2^2(ha)}{K_2^2(qa)}\bigg\{ \frac{\beta^2}{2q^2}\bigg[ (1-u)^2[K_0(qa)K_2(qa)-K_1^2(qa)]\nn\\
&  +(1+u)^2[K_2(qa)K_4(qa)-K_3^2(qa)]\bigg]\nn\\
&  -K_2^2(qa)+K_1(qa)K_3(qa) \bigg\}.
\end{align}

\subsection{TE$_{01}$ mode}
The electric field components outside the fibre core for the TE$_{01}$ mode  are given by
\begin{align}
E_{r}&=0,\nn\\
E_{\varphi}&=-\frac{i}{\sqrt{\pi}qa^2}\frac{1}{\sqrt{n_1^2P_1+n_2^2P_2}}K_{1}(qr),\nn\\
E_{z}&=0,\label{ecircular_TE01}
\end{align}
where
\begin{align} 
P_1&= \frac{1}{a^2h^2}\frac{K_0^2(qa)}{J_0^2(ha)} \left(J_1^{2}(ha)-J_0(ha)J_2(ha)\right), \nn\\
P_2 &= \frac{1}{a^2q^2} \left(K_0(qa)K_2(qa)-K_1^2(qa)\right).
\end{align}

\subsection{TM$_{01}$ mode}
The electric field components outside the fibre core for the TM$_{01}$ mode  are given by
\begin{align} 
E_{r}&=\frac{i\beta}{\sqrt{\pi}qa}\sqrt{n_1^2Q_1+n_2^2Q_2}K_1(qr),\nn\\
E_{\varphi}&=0,\nn\\
E_{z}&=\frac{1}{\sqrt{\pi}a}\sqrt{n_1^2Q_1+n_2^2Q_2}K_0(qr),\label{ecircular_TM01}
\end{align}
where
\begin{align}
 Q_1&= \frac{K_0^2(qa)}{J_0^2(ha)}\left[J_0^2(ha)+\frac{n_1^2k_{0}^{2}}{h^2}J_1^2(ha)-\frac{\beta^2}{h^2}J_0(ha)J_2(ha)\right], \nn\\
Q_2 &= \frac{\beta^2}{q^2}K_0(qa)K_2(qa)-K_0^2(qa)-\frac{n_2^2k_{0}^{2}}{q^2}K_1^2(qa).
\end{align}


\begin{thebibliography}{}

\bibitem{lewenstein_adv_phys} M.~Lewenstein, A.~Sanpera, V.~Ahufinger, B.~Damski, A.~Sen (De) and U.~Sen, Adv.~Phys.~{\bf 56}, 243 (2007).

\bibitem{bloch_rmp} I.~Bloch, J.~Dalibard and W.~Zwerger, Rev.~Mod.~Phys.~{\bf 80}, 885 (2008).

\bibitem{abrikosov} A. A. Abrikosov, Sov. Phys. JETP {\bf 5}, 1174 (1957).

\bibitem{gauge} K. J. Gunter, M. Cheneau, T. Yefsah, S. P. Rath, and J. Dalibard, Phys. Rev. A  {\bf 79}, 011604 (2009); I. B. Spielman, Phys. Rev. A {\bf 79}, 063613 (2009).

\bibitem{coddington} I. Coddington, P. Engels, V. Schweikhard, and E. A. Cornell, Phys. Rev. Lett. {\bf 91},100402 (2003)

\bibitem{spielman} Y. J. Lin, R. L. Compton, A. R. Perry, W. D. Phillips, J. V. Porto, and I. B. Spielman, Phys. Rev. Lett. {\bf 102}, 130401 (2009) ;  Y-J. Lin, R. L. Compton, K. J. Garcia, J. V. Porto and I. B. Spielman, Nature {\bf 462}, 628, (2009)

\bibitem{He_book} R. J. Donnelly, Quantized Vortices in Helium II (Cambridge University Press, 1991); 

\bibitem{He_vortex_rings}  G. W. Rayfield and F. Reif,  Phys. Rev. {\bf 136}, A1194 (1964) ; G. Gamota, Phys. Rev. Lett. {\bf 31}, 517 (1973) 

\bibitem{BEC_VR1} B. P. Anderson, P. C. Haljan, C. A. Regal, D. L. Feder, L. A. Collins, C. W. Clark, and E. A. Cornell, Phys. Rev. Lett. {\bf 86}, 2926 (2001)

\bibitem{BEC_VR2} I. Shomroni, E. Lahoud, S. Levy, and J. Steinhauer,  Nat. Phys. {\bf 5}, 193 (2009).

\bibitem{BEC_VR3}  J. Ruostekoski and Z. Dutton, Phys. Rev. A {\bf 72}, 063626 (2005) .

\bibitem{BEC_VR4} N. S. Ginsberg, J. Brand, and L. V. Hau, Phys. Rev. Lett. {\bf 94}, 040403 (2005).

\bibitem{VR_zwierlein} M. J. H. Ku, B. Mukherjee, T. Yefsah, and M. W. Zwierlein, Phys. Rev. Lett. {\bf 116}, 045304 (2016). 

\bibitem{BEC_VR5} B. Jackson, J. F. McCann, and C. S. Adams, Phys. Rev. A {\bf 61}, 013604 (1999).

\bibitem{BEC_VR6}  F. Pinsker, N. G. Berloff, and V. M. P\'{e}rez-Garc\'{i}a,  Phys. Rev. A {\bf 87}, 053624 (2013).

\bibitem{BEC_VR7}  J. Ruostekoski and J. R. Anglin,  Phys. Rev. Lett. {\bf 86}, 3934 (2001).

\bibitem{Vetsch}  E. Vetsch, D. Reitz, G. Sagu\'{e}, R. Schmidt, S. T. Dawkins, and A. Rauschenbeutel, Phys. Rev. Lett. {\bf 104}, 203603 (2010). 

\bibitem{Yariv} A. Yariv, Optical Electronics, $3^{rd}$ ed. (CBS College, New York 1985).

\bibitem{Nieddu} T. Nieddu, V. Gokhroo, and S. Nic Chormaic, J. Opt. {\bf 18}, 053001 (2016).

\bibitem{Hakuta_qdots} R. Yalla, F. L. Kien, M. Morinaga, and K. Hakuta, Phys. Rev. Lett. {\bf 109}, 063602 (2012).

\bibitem{Dowling} J. P. Dowling and J. Gea-Banacloche, Adv. At. Mol. Opt. Phys. {\bf 37}, 1 (1996).

\bibitem{FLKien1} F. L. Kien, V. I. Balykin, and K. Hakuta, Phys. Rev. A {\bf 70}, 063403 (2004). 

\bibitem{balykin2004} V. I. Balykin, K. Hakuta, F. L. Kien, J. Q. Liang, and M. Morinaga, Phys. Rev. A {\bf 70}, 011401(R) (2004).

\bibitem{helical} D. Reitz, and A. Rauschenbeutel, Optics Comm. {\bf 285}, 4705 (2012).

\bibitem{tara} C. F. Phelan, T. Hennessy, and Th. Busch, Optics Express {\bf 21}, 027093 (2013).

\bibitem{sile1} J. M. Ward, D. G. O' Shea, B. J. Shortt, M. J. Morrissey, K. Deasy, and S. Nic Chormaic, Rev. Sci. Instrumm. {\bf 77}, 083105 (2006).

\bibitem{rauschenbeutel_2008njp} G. Sagu\'{e}, A. Baade, and A. Rauschenbeutel, New J. Phys. {\bf 10}, 113008 (2008).

\bibitem{sile2} R. Kumar, V. Gokhroo, K. Deasy, A. Maimaiti, M. C. Frawley, C. Phelan, and S. Nic Chormaic, New J. Phys.  {\bf 17}, 013026 (2015).

\bibitem{sile3} M. Daly, V. G. Truong, C. F. Phelan, K. Deasy and  S. Nic Chormaic, New J. Phys. {\bf 16}, 053052 (2014).

 \bibitem{Dalibard_RMP} J. Dalibard, F. Gerbier, G. Juzeliunas, and P. Ohberg, Rev. Mod. Phys. {\bf 83}, 1523 (2011).

\bibitem{Ev_AGF} M. Mochol and K. Sacha, Scientific Reports, {\bf 5}, 7672 (2015) 

\bibitem{Ev_AGF2} V. E. Lembessis, J. Opt. Soc. Am. B. {\bf 31}, 1322 (2014). 

\bibitem{CCT:2011} C.~Cohen-Tannoudji and D.~Gury-Odelin, Advances in Atomic Physics: An Overview (World Scientific, 2011).

\bibitem{mazur_2003} L. Tong, R. R. Gattass, J. B. Ashcom, S. He, J. Lou, M. Shen, I. Maxwell, and E. Mazur, Nature {\bf 426}, 816 (2003).

\bibitem{James:17} J.~Schloss, R.~Sachdeva, and Th.~Busch, in preparation.

\end{thebibliography}
\end{document}